\documentclass[11pt]{JHEP3} 
\pdfoutput=1

\JHEPspecialurl{http://jhep.sissa.it/JOURNAL/JHEP3.tar.gz}

\usepackage{epsfig,multicol,bbm}
\usepackage{amsmath}

\title{Pulsating Strings in Lunin-Maldacena Backgrounds}

\author{Sergio Giardino and Victor O. Rivelles \\ Instituto de F\'{i}sica, Universidade de S\~ao Paulo, C. Postal 66318, 05314-970 S\~ao Paulo, SP, Brazil \\ 
	E-mail: \email{jardino,rivelles@fma.if.usp.br}}

\abstract{We consider pulsating strings in Lunin-Maldacena backgrounds, specifically in deformed Minkowski spacetime and deformed $AdS_5 \times S^5$. We find the relation between the energy and the oscillation number of the pulsating string when the deformation is small. Since the oscillation number is an adiabatic invariant it can be used to explore the regime of highly excited string states. We then quantize the string and look for such a sector. For the deformed Minkowski background we find a precise match with the classical results if the oscillation number is quantized as an even number. For the deformed $AdS_5 \times S^5$ we find a contribution which depends on the deformation parameter. }

\begin{document} 

\section{Introduction}

Integrability is an important tool for the understanding of several aspects of the AdS/CFT correspondence \cite{Dorey:2009zz}. The energy of strings in $AdS_5 \times S^5$ and anomalous dimensions of gauge invariant operators in planar ${\cal N}=4$ super Yang-Mills theory have been successfully compared with the predictions of the thermodynamic Bethe ansatz in several limits \cite{Gromov:2009zza}. Being functions of the string tension and charges these quantities usually have intricate expressions. Strings at the semiclassical level correspond to operators with large charges in the field theory and a reasonable understanding of this situation has been achieved \cite{Frolov:2002av,Beisert:2006ez}. On the other side, when the semiclassical parameters are small, corresponding to short operators in the field theory,  results for folded strings \cite{Tirziu:2008fk,Roiban:2009aa} and pulsating strings \cite{Beccaria:2010zn} are providing new data to support the correspondence. 

Despite the huge progress got so far in the $AdS_5 \times S^5$ case it is also important  to understand situations with less supersymmetry. One such a case is the planar gauge theory with a beta deformed superpotential leading to a marginally conformal ${\cal N}=2$ supersymmetric gauge theory \cite{Leigh:1995ep,Mauri:2005pa}. The dual gravitational background was found by Lunin and Maldacena \cite{Lunin:2005jy} and in the case of $AdS_5 \times S^5$ it consists of a deformation of $S^5$ through a real parameter ${\hat{\gamma}}$ which keeps the anti-de Sitter part intact while the dilaton and some RR and NS-NS fields are turned on. Both, string and gauge theories, also show signs of integrability \cite{Roiban:2003dw,Berenstein:2004ys,Frolov:2005ty,Frolov:2005dj,Beisert:2005if} with the corresponding Y-system being discussed recently \cite{Gromov:2009tv,Gromov:2010dy}. Several spinning and rotating string configurations were considered in Lunin-Maldacena backgrounds. However one important class of strings, pulsating strings, have not received much attention so far.  They have been studied in $AdS_5 \times S^5$ \cite{deVega:1994yz,Gubser:2002tv,Minahan:2002rc,Engquist:2003rn,Khan:2003sm,Arutyunov:2003za,Kruczenski:2004cn,Smedback:1998yn}, $AdS_4 \times CP^3$ \cite{Chen:2008qq,Dimov:2009rd} and other backgrounds \cite{Dimov:2004xi},\cite{Bobev:2004id},\cite{Arnaudov:2010by},\cite{Arnaudov:2010dk}. In the context of deformed geometry it was found that pulsating and rotating strings in $AdS_5 \times S^5$ are images of point like strings living in the deformed space \cite{Frolov:2005iq}.  In this paper we will provide some results for pulsating strings in deformed backgrounds when the deformation is small.  

Since the classical motion of a pulsating string is periodic we can use its oscillation number\footnote{Here $p$ is the canonical momentum to the oscillating coordinate $q$.} $N = \oint p dq/2\pi$ and its energy to characterize the dynamics. The string oscillation number is not one of the string charges but it is very useful to  describe the behavior of pulsating strings as shown in the $AdS_5 \times S^5$ case \cite{Kruczenski:2004cn,Beccaria:2010zn}. So we consider pulsating strings in deformed Minkowski spacetime and $AdS_5 \times S^5_{\hat{\gamma}}$ to find the relation between the energy and the oscillation number in the regime of small deformation. Since the oscillation number is an adiabatic invariant it can provide some information about the semi-classical regime where $N$ is very large.
We then perform the quantization of the string looking for highly excited string states and compute its energy in perturbation theory. When we consider the deformed Minkowski spacetime we find full agreement if the oscillation number is quantized as an even number. In the $AdS_5 \times S^5_{\hat{\gamma}}$ case we consider only strings oscillating in the deformed sphere since $AdS_5$ is not deformed and the solutions in this sector would be the same as in the undeformed case. We find that the oscillation number can be expressed in terms of elliptic functions. We derive the relation between the energy and the oscillation number for short strings in the low energy regime. When the deformation vanishes we recover the results for $AdS_5 \times S^5$ \cite{Beccaria:2010zn}. In the case of large energy we consider the quantization of highly excited strings up to second order in perturbation theory and we find that the energy has a new term proportional to the deformation parameter which is not present in the classical relation between the energy and the oscillation number. We also discuss the contributions coming from the fluctuations of the radial $AdS$ coordinate.

This paper is organized as follows. A short discussion of the Lunin-Maldacena backgrounds is presented in the next section. In Section 3 we consider pulsating strings in the deformed Minkowski spacetime and in Section 4 we consider the case of $AdS_5 \times S^5_{\hat{\gamma}}$. Finally in Section 5 we present some conclusions.

\section{Lunin-Maldacena Backgrounds}

The beta deformed ${\cal N}=4$ SYM gauge theory has a $U(1)\times U(1)$ global symmetry which is realized geometrically as an isometry of a two torus in the dual string background. It can also be understood as a TsT transformation applied to $AdS_5\times S^5$  and can be generalized to other backgrounds as well \cite{Frolov:2005dj,Alday:2005ww}. Keeping with the standard notation the real deformation parameter in the string side is denoted by ${\gamma}$. Then the Lunin-Maldacena background is given by \cite{Lunin:2005jy}
\begin{eqnarray}
ds^{2}&=&R^{2}\left[ds_{AdS_{5}}^{2}+\sum_{i=1}^{3}(d\mu_{i}^{2}+G\mu_{i}^{2}d\phi_{i})+\hat{\gamma}^{2}\mu_{1}^{2}\mu_{2}^{2}\mu_{3}^{2}{G\left(\sum_{i=1}^{3}d\phi_{i}\right)}^{2}\right], \label{AdS5 x S5 def}\\
B_{2}&=&\hat{\gamma}R^{2}G\,\left(\mu_{1}^{2}\mu_{2}^{2}\,
  d\phi_{1}\wedge d\phi_{2}+\mu_{2}^{2}\mu_{3}^{2}\, d\phi_{2}\wedge
  d\phi_{3}-\mu_{1}^{2}\mu_{3}^{2}\, d\phi_{1}\wedge
  d\phi_{3}\right), \nonumber
\\e^{2\Phi}&=&e^{2\Phi_{0}}G, \nonumber\\
C_{2}&=&-48\pi\, N\,\hat{\gamma}\,\omega_{1}\wedge d\psi, \nonumber \\
C_{4}&=&16\pi\, N\,(\omega_{4}+G\, d\omega_{1}\wedge d\phi_{1}\wedge d\phi_{2}\wedge
d\phi_{3}), \nonumber
\end{eqnarray}
where $B_2$ is the NS-NS two-form potential, $C_2$ and $C_4$ are the two and four-form RR potentials respectively, $\Phi$ is the dilaton, and 
\begin{eqnarray}
&&ds_{AdS_5}^{2}=-\cosh^2\rho\,dt^2+d\rho^2+\sinh^2\rho\left(d\Psi^2+\sin^2\Psi
  d\Phi_1^2+\cos^2\Psi d\Phi_2^2\right),\label{ads_sec}\\
&& G^{-1}=1+\hat{\gamma}^{2}\left(\mu_{1}^{2}\mu_{2}^{2}+\mu_{2}^{2}\mu_{3}^{2}+\mu_{1}^{2}\mu_{3}^{2}\right),\\
&& {d\omega}_{1}=\sin^{3}\alpha\cos\alpha\,\sin\theta\cos\theta
d\alpha\wedge d\theta, \qquad {\omega}_{AdS_{5}}=d\omega_{4}, \\
&& \sum_{i=1}^{3}\mu_{i}^{2}=1,\qquad\hat{\gamma}=R^{2}\gamma, \qquad R^{4}=4\pi e^{\Phi_{0}}N, \label{c1}
\end{eqnarray}
with $R$ being the AdS radius. 
Notice that only the $S^5$ sphere has been modified and that 
when ${\hat{\gamma}}=0$ we recover the original $AdS_{5}\times S^{5}$ background. 

We can solve the constraint in (\ref{c1}) by introducing spherical like coordinates 
\begin{eqnarray}
\mu_{1}&=&\sin\theta\cos\psi, \nonumber\\
\mu_{2}&=&\cos\theta\label{mu i}, \\
\mu_{3}&=&\sin\theta\sin\psi,\nonumber
\end{eqnarray}
so that the metric and the KR field become
\begin{eqnarray}
ds^{2}&=&R^{2}\left\{ ds_{AdS_5}^{2}+d\theta^{2}+\sin^2\theta
  d\psi^{2}+\right. \nonumber\\
&&+G\left[\sin^2\theta\cos^{2}\psi\left(1+\hat{\gamma}^{2}\sin^2\theta\cos^{2}\theta\sin^{2}\psi\right)d\phi_{1}^{2}+\right.\nonumber\\
&&+\cos^{2}\theta\left(1+\hat{\gamma}^{2}\sin^{4}\theta\cos^{2}\psi\sin^{2}\psi\right)d\phi_{2}^{2}+\nonumber\\ 
&&+\left.\sin^2\theta\sin^{2}\psi\left(1+\hat{\gamma}^{2}\sin^2\theta\cos^{2}\theta\cos^{2}\psi\right)d\phi_{3}^{2}\right]+\nonumber\\
&&\left.+2G\hat{\gamma}^2\sin^{4}\theta\cos^{2}\theta\sin^2\psi\cos^{2}\psi\left(d\phi_{1}d\phi_{2}+d\phi_{2}d\phi_{3}+d\phi_{1}d\phi_{3}\right)\right\}, \label{LM metrica 2} \\
B_{2}&=&R^{2}\hat{\gamma}G\sin^2\theta\left(\cos^{2}\theta\cos^{2}\psi
  d\phi_{1}\wedge d\phi_{2}+\cos^{2}\theta\sin^2\psi
  d\phi_{2}\wedge d\phi_{3}-\right.\nonumber\\ &&\left.-\sin^2\theta\sin^2\psi\cos^{2}\psi d\phi_{1}\wedge d\phi_{3}\right), 
\end{eqnarray}
where now 
\begin{equation}
G^{-1}=1+\hat{\gamma}^2\sin^{2}\theta(\cos^{2}\theta+\sin^2\theta\cos^{2}\psi\sin^2\psi).
\end{equation}
We will use this form to study pulsating strings. 

The Lunin-Maldacena deformation technique can also be applied to the ten
dimensional Minkowski spacetime \cite{Lunin:2005jy}. The result is simpler than in the $AdS_5 \times S^5$ case and it will be used in the next section as a background for pulsating strings even though no dual gauge theory is known. The ten dimensional Minkowski spacetime is split into a four dimensional one and the remaining six dimensional space is deformed resulting in 
\begin{eqnarray}
ds^{2}&=&\eta_{\mu\nu}dx^{\mu}dx^{\nu}+\sum_{i=1}^{3}\left(dr_{i}^{2}+Gr_{i}^{2}d\phi_{i}^{2}\right)+\gamma^{2}r_{1}^{2}r_{2}^{2}r_{3}^{2}G\left(\sum_{i=1}^{3}d\phi_{i}\right)^{2}, \label{R10
deformado}\\
B_{2}&=& \gamma 
G\left(r_{1}^{2}r_{2}^{2}\, d\phi_{1}\wedge
  d\phi_{2}+r_{2}^{2}r_{3}^{2}\, d\phi_{2}\wedge
  d\phi_{3}-r_{1}^{2}r_{3}^{2}\, d\phi_{1}\wedge
  d\phi_{3}\right), \nonumber\\
  e^{2\Phi}&=&G,\qquad
G^{-1}=1+\gamma^{2}\left(r_1^2r_2^2+r_2^2r_3^2+r_1^2r_3^2\right).\nonumber
\end{eqnarray}
No RR fields are present in this case. As expected, the limit $\gamma=0$ reproduces the ten dimensional Minkowski spacetime.

\section{Pulsating Strings in Deformed Minkowski Spacetime}

To consider pulsating strings in (\ref{R10 deformado}) we will use spherical coordinates defined by
\begin{align}
	r_1 = r \sin\theta\cos\psi,  \qquad r_2 = r \sin\theta \sin\psi, \qquad r_3 = r \cos\theta. 
\end{align}
Firstly we choose a string at the origin of the Minkowski spacetime and with  $\psi=\pi/2$, $\phi_1 = \phi_{2}=0$ and  $\theta$ fixed so that the metric becomes
\begin{eqnarray}
ds^2&=&-dt^2+dr^2 + G r^2 \cos^2\theta d\phi_3^2, \label{metr1}\\
G^{-1}&=&1+\gamma^{2}r^{4}\sin^2\theta\cos^2\theta.\label{(G-1)}
\end{eqnarray}
With this choice the coupling of the string to the $B_2$ field vanishes.
We now consider an ansatz for a pulsating string that is wound in $\phi_3$ and oscillating in the radial direction 
\begin{equation}
t=\kappa\tau, \qquad r=r(\tau), \qquad \phi_3=m\sigma, \qquad \theta = \mbox{constant},
\label{ansatz minahan R}
\end{equation}
with $m$ being the winding number.  
The only non trivial equation comes from the Virasoro constraint 
\begin{equation}
\dot{r}^{2} + m^{2}r^{2}G\cos^2\theta= \kappa^2. 
\label{vinc r def}
\end{equation}
At this point we must notice that the fixed angle $\theta$ appears always in the combination  $m \cos\theta$ and $\gamma \sin\theta \cos\theta$ which allow us to rescale $m$ and $\gamma$ as  $m \cos\theta \rightarrow m$ and $ \gamma \sin\theta \cos \theta \rightarrow \gamma$ respectively, assuming that $\theta$ is different from zero and $\pi/2$. 

\begin{figure}[h] 
       \centering  
       \includegraphics[height=4cm]{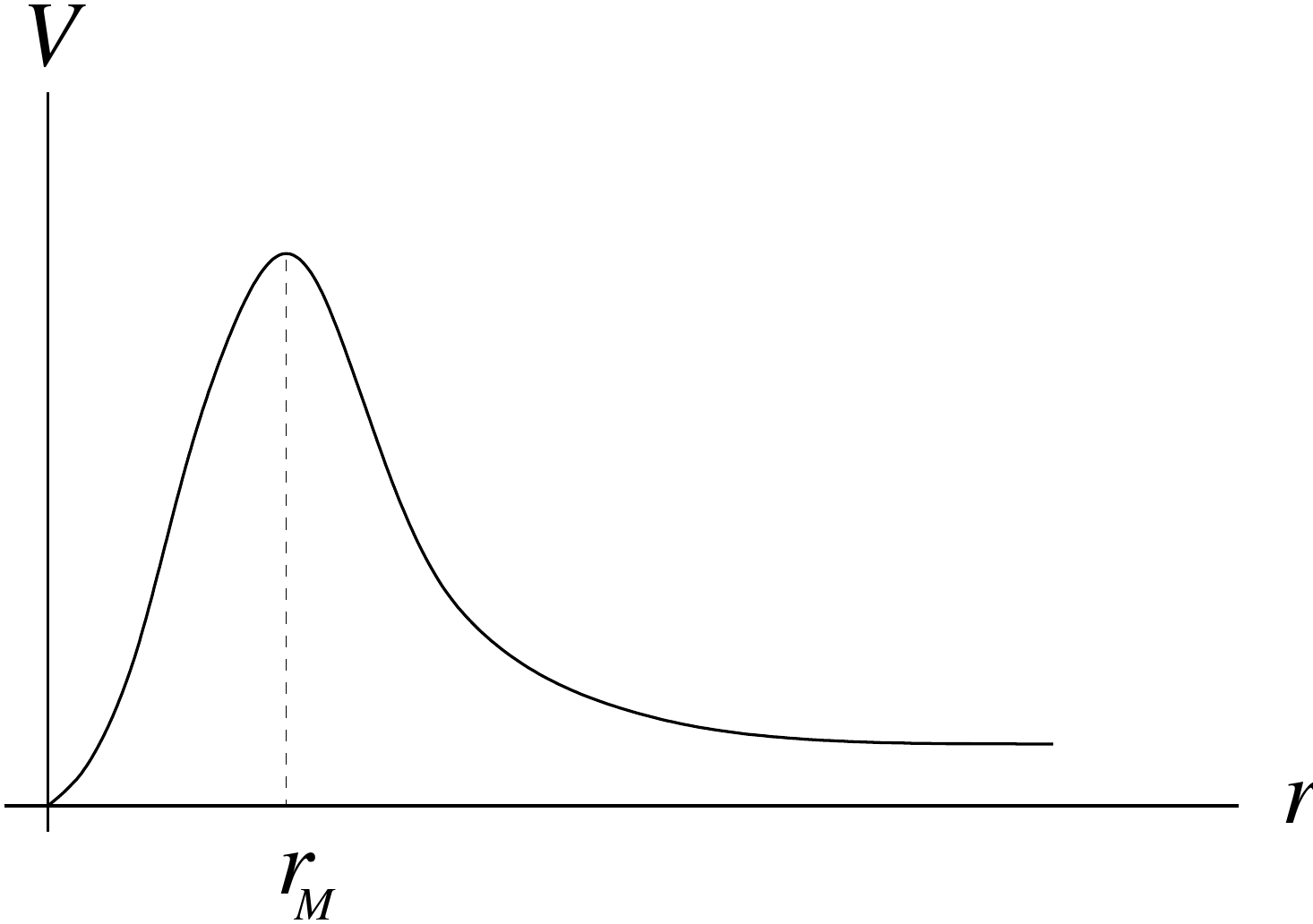}
       \caption{\it The effective potential in deformed Minkowski spacetime}\label{figA}
\end{figure}

We can write (\ref{vinc r def}) in terms of an effective potential as 
\begin{equation}
\frac{\dot{r}^2}{m^2} = \frac{\kappa^2}{m^2} - V(r). \label{r ponto}
\end{equation}
where $V(r)$ is given by
\begin{equation}
V(r)=\frac{ r^2}{1+\gamma^{2}r^{4}},
\label{V(r)}
\end{equation}
The behavior of the effective potential is depicted in Fig. \ref{figA}. 
There is an unstable equilibrium point at $r_{M}^2=1/\gamma$  where the potential reaches its maximum value $V(r_{M})=1/(2\gamma)$. 
The points where the radial velocity vanishes are 
\begin{equation}
r{}_{\pm}^{2}=\frac{1}{2\gamma^2} \frac{m^2}{\kappa^2}\left( 1 \pm \sqrt{ 1- 4\gamma^2 \frac{\kappa^4 }{m^4}} \right), \label{r+-}
\end{equation}
and the condition for oscillatory motion implies that $\kappa^2/m^2 \le 1/(2\gamma)$. 

Now (\ref{vinc r def}) can be rewritten as
\begin{equation}
dr \, \sqrt{\frac{1 + \gamma^2 r^4 }{\left(r^2-r_{+}^{2}\right)\left(r^{2}-r_{-}^{2}\right)}}= \gamma \, \kappa \, d\tau.
\label{eq dif r}
\end{equation}
Let us consider the case of small deformation $\gamma<<1$. Then the height of the potential is very large and we can consider a highly excited pulsating string. Its radial motion will be limited to $r_-$. For small $\gamma$ the turning points are 
\begin{eqnarray}
r_{+}^{2}&\approx& \frac{1}{\gamma^2} \frac{m^{2}}{\kappa^2} \left( 1 - \gamma^2 \frac{\kappa^4}{m^4} \right), \label{r+_pequeno}\\
r_{-}^{2}&\approx& \frac{\kappa^2}{m^{2}} \left( 1 +  \gamma^2 \frac{\kappa^4}{m^{4}} \right). \label{r-_pequeno}
\end{eqnarray}
In this situation $\gamma^2 r^4 <1$ and the square root in (\ref{eq dif r}) can be expanded up to first order in $\gamma^2$ resulting in 
\begin{align}
\intop_{0}^{r}dr\frac{1 + \gamma^2 r^4/2} {\sqrt{\left(r^2-r_{+}^{2}\right)\left(r^{2}-r_{-}^{2}\right)}}
= {\gamma\,\kappa}(\tau - \tau_{0}),
\label{eq dir r expand}
\end{align}
where $\tau_0$ is an integration constant. The integrals can be written in terms of elliptic functions. 
%
%
%
Notice that for vanishing ${\gamma}^2$ we recover the results for flat spacetime discussed for instance in \cite{Minahan:2002rc}. 

In order to characterize the string dynamics we can compute the relation between the string oscillation number and the energy. The oscillation number associated to the periodic radial motion is $N = \oint \Pi_r dr/2\pi$, where $\Pi_r$ is the momentum conjugated to $r$. To compute $\Pi_r$ we take the Polyakov action for the ansatz (\ref{ansatz minahan R})
\begin{equation}
S=-\frac{\sqrt{\lambda} \kappa}{2}\int d\tau \left( 1 - \dot{r}^2 + m^2 r^2 G \right), 
\end{equation} 
to find that $\Pi_r = \sqrt{\lambda} \kappa \dot{r}$. Then 
\begin{align}
	N &= \frac{\sqrt{\lambda} \kappa}{2\pi} \oint \dot{r} dr = \frac{\sqrt{\lambda} \kappa}{2\pi} \oint \sqrt{\kappa^2 - m^2 V(r)} dr \\
	&= 2\frac{\sqrt{\lambda}}{\pi}  \gamma \kappa  \int_0^{r_-} dr \sqrt{(r^2 - r_+^2)(r^2 - r_-^2)}( 1 - \frac{1}{2} \gamma^2 r^4 ),
\end{align}
where in the last line we used the small $\gamma$ limit. The integrals can be easily evaluated in the limit of small $\gamma$  
and we get
\begin{equation}
{N}= \frac{1}{2} \sqrt{\lambda} \frac{\kappa^2}{m} \left( 1 +   \frac{5}{16} \gamma^2 \frac{\kappa^4}{m^4} \right).
\end{equation}
We can then solve for the energy $E = \sqrt{\lambda} \kappa$ in the limit of small $\gamma$ to find
\begin{equation} \label{energy}
	E =  \sqrt{2 \lambda^{1/2} m N} \left( 1 - \frac{5}{8} \frac{\gamma^2}{\lambda} \frac{N^2}{m^2}  \right).
\end{equation}
Since adiabatic invariants can be used to probe the highly excited states of the quantum theory the above expression can be regarded an approximation for the energy in the limit of large quantum numbers in the radial direction. In what follows we will check that this is indeed true. 

Quantization will be performed by starting with the Nambu-Goto action for the ansatz (\ref{ansatz minahan R})
\begin{align}
	S = - m \frac{\sqrt{\lambda}}{\kappa} \int dt \,  r \, \sqrt{G} \sqrt{1 - \dot{r}^2}.
\end{align}
We find that $\Pi_r = \sqrt{\lambda} m r \sqrt{G} \dot{r}/ \sqrt{k^2 - \dot{r}^2}$ and 
\begin{eqnarray}
H^2 &=& \Pi_r^{2} + \lambda m^2 \frac{r^2}{1 + \gamma^2 r^4} \\
    &=& \Pi_r^{2} + \lambda m^2 r^2  - \gamma^2 \lambda m^2 \frac{r^6}{1 + \gamma^2 r^4}, \label{Hamilt Min} 
\end{eqnarray}
where in the last line we split the potential in a term which is independent of $\gamma$ and another which depends on the deformation. 
If the deformation vanishes we get a radial harmonic oscillator potential $\lambda m^2 r^2$. 
In order to proceed we assume that the wave function depends only on $r$ so we have to realize $\Pi_r^2$ as the radial component of the Laplacian
\begin{align}\label{lap}
	\Pi_r^2 = - \frac{1}{\sqrt{-g}} \frac{d}{dr} \left( \sqrt{-g} \frac{d}{dr} \right).
\end{align}
We will first consider the situation where we quantize only the radial motion on the deformed plane (\ref{metr1}) ignoring the remaining coordinates, that is we take $\sqrt{-g} = r \cos\theta \sqrt{G}$ in (\ref{lap}). In this situation we expect to reproduce (\ref{energy}) for higher quantum numbers. The Schr\"{o}dinger equation to be solved is then 
\begin{align}\label{H1}
	H^2 \Psi = - \frac{1}{r} \frac{d}{dr} \left( r \frac{d\Psi}{dr} \right) + \lambda m r^2 \Psi + 2 \gamma^2 r^3 \frac{d\Psi}{dr} - \gamma^2 \lambda m^2 \frac{r^6}{1 + \gamma^2 r^4} \Psi = E^2 \Psi,
\end{align}
where we took the limit of small deformation in the third term. To solve (\ref{H1}) we will use standard perturbation theory. 
First we have to choose the unperturbed Hamiltonian. We can either choose the two first terms or the three first terms. In the last case we are considering the quantization in the deformed space while in the first one we are regarding the deformation as a perturbation and quantizing in the undeformed space. Either way we get the same result at the end. So we choose as the unperturbed Hamiltonian the first two terms of (\ref{H1}) and find that the normalized wave function is 
\begin{equation}
	\Psi_n(r) = \sqrt{2 \lambda^{1/2} m} \,\, e^{-\frac{1}{2} \sqrt{\lambda} m r^2} L_n(\sqrt{\lambda} m r^2),
\end{equation}
where $L_{n}$ are Laguerre polynomials. For highly excited states we have
\begin{equation}
	E^2_{0,n} = 4\sqrt{\lambda} \, m \, n, \label{energy1}
\end{equation}
or $E_{0,n} = \sqrt{ 2 \lambda^{1/2} m (2n)}$. Since we are quantizing a radial harmonic oscillator we expect that its energy depends only on even integers. Comparison with the lowest order of (\ref{energy}) shows complete agreement if the oscillation number is quantized as $2n$. 

The first order correction to the energy for small deformation is 
\begin{align}\label{E2}
	\delta E^2_n &= 2 \gamma^2 \int_0^\infty dr \, r^4 \Psi_n^\star \frac{d \Psi_n}{dr} - \gamma^2 \lambda m^2 \int_0^\infty dr \, \frac{r^7}{1 + \gamma^2 r^4} \, |\Psi_n|^2 \nonumber \\
	&= 2 \frac{ \gamma^2}{\sqrt{\lambda} m} n   -  20 \frac{\gamma^2}{\sqrt{\lambda m}} n^3.
\end{align}
The first term can be disregard for large $n$ and the energy is then 
\begin{equation} \label{correct1}
	E_n = \sqrt{2\lambda^{1/2} m (2n)} \left( 1 - \frac{5}{8} \frac{\gamma^2}{\lambda} \frac{(2n)^2}{m^2} \right).
\end{equation}
As expected it agrees with (\ref{energy}) if we assume that the oscillation number $N$ is quantized with eigenvalue $2n$. 

We can capture some of the quantum effects of the deformed ten dimensional space by taking in  $\Pi_r^2$ the contribution of all dimensions by choosing $\sqrt{-g} = r^5 \sqrt{G} \dots$ where the dots represent terms which do not depend on $r$ and cancel out in (\ref{lap}). We also  assume that the wave function depends only on the radial coordinate. The Schr\"{o}dinger equation for $H^2$ reads now 
\begin{align}
	H^2 \Psi = - \frac{1}{r^5} \frac{d}{dr} \left( r^5 \frac{d \Psi}{dr} \right) + \lambda m^2 r^2 \Psi + 4 \gamma^2 r^3 \frac{d \Psi}{dr} - \gamma^2 \lambda m^2 \frac{r^6}{1 + \gamma^2 r^4} \Psi = E^2 \Psi,\label{ham_flat}
\end{align}
where in the third term we already took the limit of small deformation. 
Again we take the unperturbed potential as that of the radial harmonic oscillator and consider the last two terms as perturbations. The normalized wave function is now
\begin{align}
	\Psi_n(r) = \sqrt{2} \lambda^{3/4} \frac{m^{3/2}}{n} e^{-\sqrt{\lambda} m r^2/2} L^{(2)}_n(\sqrt{\lambda} m r^2),
\end{align}
where $L^{(2)}_n$ are generalized Laguerre polynomials. The energy for high quantum number is
still given by the former result (\ref{energy1}). This happens because only the zero point contribution to the energy changes with the dimension and in the limit of large $n$ it can be ignored. 

Now we have two corrections coming from the deformation. The first one is the $r^6/(1 + \gamma^2 r^4)$ perturbation in the potential term in (\ref{ham_flat}). For large $n$ it gives the same correction of order $n^3$ as in (\ref{E2}) even though the wave function and the number of dimensions are different. The second correction in (\ref{ham_flat}) has a derivative term and in the limit of large $n$ we get 
\begin{align}
	\delta^{'} E^2_n = - 32 \frac{\gamma^2}{\sqrt{\lambda}} \frac{n}{m}.
\end{align}
It is of order $n$ while the first term is of order $n^3$ so it can be disregarded. This means that in the large $n$ limit we get the same result (\ref{correct1}) as in the planar case. This shows that for a highly excited string the classical result (\ref{energy}) can be straightforwardly taken to quantum case when $N$ is quantized as $2n$.

\section{Pulsating Strings in the Deformed Sphere of $AdS_5 \times S^5_{\hat{\gamma}}$}

We now consider a pulsating string in the deformed sphere of $AdS_5 \times S^5_{\hat{\gamma}}$. We assume that in the $AdS_5$ sector (\ref{ads_sec}) we have $\Psi=\Phi_1=\Phi_2=0$ while in the  $S^5_{\hat{\gamma}}$ sector (\ref{LM metrica 2}) we take $\phi_1=\phi_2=0$ and $\psi=\pi/2$ so that we still have a deformed sphere inside $S^5_{\hat{\gamma}}$. In this situation the metric reduces to 
\begin{equation} \label{metr2}
ds^{2}= R^{2}\left(-\cosh^2\rho\, dt^2+d\rho^2+d\theta^2+G\sin\theta^2\, d\phi_3^{2}\right),
\end{equation}
while $B_2=C_2=0$, and now $G = 1/(1 + \hat{\gamma}^2 \sin^2\theta \cos^2\theta)$.  

\begin{figure}[h]
       \centering  
       \includegraphics[height=4cm]{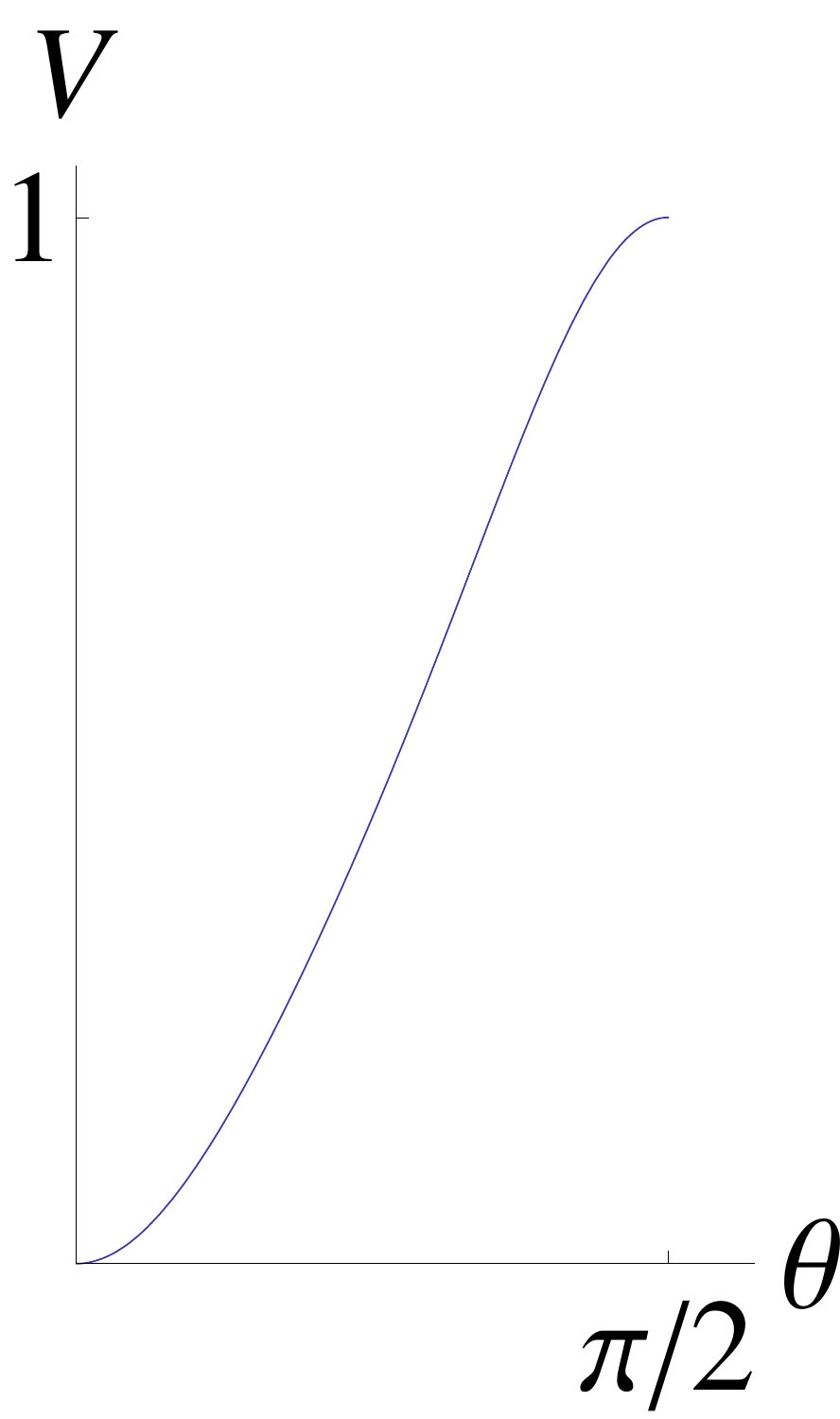}
       \caption{\it The effective potential in $AdS_5 \times S^5_{\hat{\gamma}}$}
       \label{fig4}
\end{figure}

We consider the following ansatz for a pulsating string wound $m$ times along $\phi_3$ in the deformed sphere 
\begin{equation} \label{ans2}
t = \kappa \tau,\ \qquad\rho=\rho(\tau),\qquad\theta=\theta(\tau), \qquad\phi_3=m\sigma.
\end{equation}
The equations of motion are satisfied if $\rho$ is constant, so we take the string at the center of $AdS_5$. Together with the Virasoro constraints we get the only non-trivial equation 
\begin{equation}
 \dot{\theta}^{2} - \kappa^2 + m^2G\sin^2\theta=0. 
\label{eq mov final}
\end{equation}
As before we introduce an effective potential as 
\begin{eqnarray}
\frac{\dot{\theta}^2}{m^2} &=& \frac{\kappa^2}{m^2} - V(\theta) \nonumber \\
V(\theta)&=& \frac{\sin^2\theta}{1+\hat{\gamma}^2\sin^2\theta\cos^2\theta}.
\end{eqnarray}
which is plotted in Fig.\ref{fig4}. The potential has a maximum at $\pi/2$ with value 1 so that there are two situations to be analyzed. If $\kappa^2/m^2<1$ then there is a turning point so that $\theta$ is limited to a maximum value $\theta_+$. In this case the energy is small $E^2 < \lambda m^2$ and we have a short string oscillating on the deformed sphere. If $\kappa^2/m^2>1$ then $E^2 > \lambda m^2$ and there are no turning points. We now have a string oscillating all the way from the equator to one of the poles of the deformed sphere. 

We can rewrite (\ref{eq mov final}) as
\begin{align}
	\frac{\dot{\theta}^2}{m^2} = \hat{\gamma}^2 \frac{\kappa^2}{m^2} \frac{(\sin^2\theta_+ - \sin^2\theta)(\sin^2\theta - \sin^2\theta_-)}{1 + \hat{\gamma}^2 \sin^2\theta \cos^2\theta},
\end{align}
where, in the limit of small deformation, we have 
\begin{align}
	\sin^2\theta_+ = \frac{\kappa^2}{m^2} \left[ 1 + \hat{\gamma}^2 \frac{\kappa^2}{m^2} \left( 1 - \frac{\kappa^2}{m^2} \right) \right], \qquad \sin^2\theta_- = - \frac{1}{\hat{\gamma}^2} \frac{1}{\sin^2\theta_+}.
\end{align}
For $\kappa^2/m^2<1$ there is one turning point $\theta_+$ while for $\kappa^2/m^2>1$ there is none. 

To find out the oscillating number we need the Polyakov action for the ansatz (\ref{ans2})
\begin{align}
	S = - \frac{\sqrt{\lambda} \kappa}{2} \int d\tau ( \cosh^2\rho - \dot{\rho}^2 - \dot{\theta}^2 + m^2 \sin^2\theta G), 
\end{align}
from which it follows that $\Pi_\theta = \sqrt{\lambda} \kappa \dot{\theta}$. Then the oscillating number is 
\begin{align}
	N = \frac{1}{2\pi} \oint \Pi_\theta \, d\theta = \frac{1}{2\pi} \sqrt{\lambda} \kappa \oint \dot{\theta} \, d\theta.
\end{align}

Let us consider first the case of short strings when $\kappa^2/m^2<1$. The oscillation number in this case is given by
\begin{align}
 N 
 = \frac{2\sqrt{\lambda}}{\pi} \frac{\kappa}{\sin\theta_+} \int_0^{\theta_+} d\theta \sqrt{\sin^2\theta_+ - \sin^2\theta} \left[ 1 + \frac{1}{2} \hat{\gamma}^2 ( \sin^2\theta_+ - \cos^2\theta) \sin^2\theta \right].
\end{align}
The integrals reduce to elliptic functions which still are $\hat{\gamma}$ dependent. After further expansion in $\hat{\gamma}$ we find 
\begin{align}
	N &= \frac{2\sqrt{\lambda}}{\pi} m \left[ \mathbb{E}\left(\frac{\kappa^2}{m^2}\right) - \left( 1 - \frac{\kappa^2}{m^2} \right) \mathbb{K}\left(\frac{\kappa^2}{m^2}\right) \right] \nonumber \\
	&+ \hat{\gamma}^2 \frac{\sqrt{\lambda}}{\pi} m \left[ - \frac{1}{15}\left( 2 + 3 \frac{\kappa^2}{m^2} - 8 \frac{\kappa^4}{m^4} \right) \mathbb{E} \left( \frac{\kappa^2}{m^2}\right) + \frac{2}{15} \left( 1 - \frac{\kappa^2}{m^2} \right) \left( 1 + 2 \frac{\kappa^2}{m^2} \right) \mathbb{K}\left(\frac{\kappa^2}{m^2}\right) \right],
\end{align}
where $\mathbb{K}$ and $\mathbb{E}$ are complete elliptic integrals of the first and second kind respectively. The string energy $E= \sqrt{\lambda} \kappa$ can be used to eliminate $\kappa$ and allows to find the relation between the energy $E$ and the oscillation number $N$. To do that we have to expand the elliptic functions and we have to assume that $E/(\sqrt{\lambda}m)$ is small. We then obtain 
\begin{align}
	E &= \sqrt{2 \lambda^{1/2} m N} \left[ 1 - \frac{1}{8} \frac{N}{\sqrt{\lambda} m} - \frac{5}{128} \left(\frac{N}{\sqrt{\lambda}m} \right)^2 + \dots \right] \nonumber \\
	&- \frac{3\sqrt{2}}{8} \hat{\gamma}^2 \frac{N^{3/2}}{(\sqrt{\lambda} m)^{1/2}} \left[ 1 - \frac{41}{24} \frac{N}{\sqrt{\lambda}m} + \frac{193}{384} \left( \frac{N}{\sqrt{\lambda}m} \right)^2 + \dots \right].
\end{align}
As expected there is a natural small expansion parameter $N/(\sqrt{\lambda} m)$ involving the oscillation number so we are in the regime of low quantum numbers. Also our results reduce to the $AdS_5 \times S^5$ results \cite{Beccaria:2010zn} in the undeformed case. Notice we can not perform the quantization for strings with very large energy because we had to assume that $E/(\sqrt{\lambda}m)$ is small when the elliptic functions were expanded. 

When $\kappa^2/m^2>1$ the oscillation number is
\begin{align}
	N = \frac{2 \sqrt{\lambda}}{\pi} \frac{\kappa}{\sin\theta_+} \int_0^{\pi/2} d\theta \sqrt{\sin^2\theta_+ - \sin^2\theta} \left[ 1 + \frac{1}{2} \hat{\gamma}^2 ( \sin^2\theta_+ - \cos^2\theta) \sin^2\theta \right].
\end{align}
After taking the limit of small deformation and expanding the elliptic functions we get 
\begin{align}
	N &= 2 \frac{\sqrt{\lambda}\kappa}{\pi} \mathbb{E} \left( \frac{m^2}{\kappa^2} \right) + \frac{1}{15} \hat{\gamma}^2 \frac{\sqrt{\lambda} \kappa}{\pi} \left[ - 2 \left( 1 + \frac{3}{2} \frac{\kappa^2}{m^2} - 4 \frac{\kappa^4}{m^4} \right) \mathbb{E} \left( \frac{m^2}{\kappa^2} \right) \right. \nonumber \\
	  & \left.+ \left( 1 - \frac{\kappa^2}{m^2} \right) \left( 1 + 8 \frac{\kappa^2}{m^2} \right) \mathbb{K} \left( \frac{m^2}{\kappa^2} \right) \right].
\end{align}
The elliptic functions can be expanded in order to express the energy in terms of the oscillation number. In order to do that we have to assume that $E/(\sqrt{\lambda}m)$ is large and we obtain
\begin{align} \label{class2}
	 E &= N \left[ 1 + \frac{1}{4} \left( \frac{\sqrt{\lambda}m}{N} \right)^2 - \frac{1}{64} \left( \frac{\sqrt{\lambda}m}{N} \right)^4 + \dots \right] \nonumber \\
	 &- \frac{1}{32} \hat{\gamma}^2 \lambda \frac{m^2}{N} \left[ 1 - \frac{3}{16} \left( \frac{\sqrt{\lambda}m}{N} \right)^2 + \frac{5}{128} \left( \frac{\sqrt{\lambda}m}{N} \right)^4 + \dots \right].
\end{align}
Now the small expansion parameter is $\sqrt{\lambda}m/{N}$ so we can consider the regime of large quantum numbers. 

To quantize the string we consider the Nambu-Goto action for the ansatz (\ref{ans2}). Going to the Hamiltonian formalism we get 
\begin{align}
	\Pi_\rho &= \frac{\sqrt{\lambda} m \sin\theta \sqrt{G}}{\sqrt{\cosh^2\rho - \dot{\rho}^2 - \dot{\theta}^2}} \dot{\rho},  \\
	\Pi_\theta &= \frac{\sqrt{\lambda} m \sin\theta \sqrt{G}}{\sqrt{\cosh^2\rho - \dot{\rho}^2 - \dot{\theta}^2}} \dot{\theta},
\end{align}
and
\begin{align}
	H^2 &= \cosh^2\rho \,\, ( \Pi_\rho^2 + \Pi_\theta^2 + \lambda m^2 G \sin^2\theta), \nonumber \\ 
	 &= \cosh^2\rho \,\, [ \Pi_\rho^2 + \Pi_\theta^2 + \lambda m^2  \sin^2\theta( 1 - \hat{\gamma}^2 \sin^2\theta \cos^2\theta)] \label{HAM},
\end{align}
where in the last line we took the limit of small deformation. For the moment let us consider the string sitting at the center of $AdS$. Firstly we will quantize the string in the deformed sphere (\ref{metr2}). Then $\Pi_\theta^2$ has to be realized like (\ref{lap}) (with $r$ replaced by $\theta$) using $\sqrt{-g} = \sqrt{G} \sin\theta $.  For small deformation the Schr\"{o}dinger equation reads
\begin{align} \label{ham2}
	H^2 \Psi &= - \frac{1}{\sin\theta} \frac{d}{d\theta} \left( \sin\theta \frac{d\Psi}{d\theta} \right) + \hat{\gamma}^2 \sin\theta \cos\theta ( 1 - 2 \sin^2\theta ) \frac{d\Psi}{d\theta} \nonumber \\
	&+ \lambda m^2 \sin^2\theta ( 1 - \hat{\gamma}^2 \sin^2\theta \cos^2\theta ) \Psi = E^2 \Psi.
\end{align}
Taking the first term as the unperturbed Hamiltonian the normalized eigenfunctions are written in terms of Legendre polynomials
\begin{align} 
	\Psi_n(\theta) = \sqrt{2n+1} P_n(\cos\theta),
\end{align}
with eigenvalue $E^2_{0,n} = n(n+1)$. For highly excited strings $E_{0,n} = n$ and it agrees with lowest order term in (\ref{class2}) if the oscillation number is quantized as an integer. 

The first order correction to the energy is then computed in perturbation theory. For highly excited states we find
\begin{align} \label{correc1}
	&\delta_1 E^2_n = \nonumber \\
	&= \int_0^{\pi/2} d\theta \, \sin\theta \, \Psi^\star_n \left[  \hat{\gamma}^2 \sin\theta \cos\theta ( 1 - 2 \sin^2\theta ) \frac{d}{d\theta} + \lambda m^2 \sin^2\theta ( 1 - \hat{\gamma}^2 \sin^2\theta \cos^2\theta ) \right]  \Psi_n \nonumber \\ 
&= \frac{1}{2} \lambda m^2 \left( 1 - \frac{1}{8} \hat{\gamma}^2 \right).
\end{align}
The second order correction can also be computed in the large $n$ limit and we get 
\begin{align} \label{correc2}
	\delta_2 E^2_n = \frac{1}{32} \frac{\lambda^2 m^4}{n^2} \left( 1  - \frac{1}{8} \hat{\gamma^2} \right).
\end{align}
We then find for the energy of highly excited string states
\begin{align} \label{fullenergy}
	E_n = n \left[ 1 + \frac{1}{4} \frac{\lambda m^2}{n^2} - \frac{1}{64} \left( \frac{\lambda m^2}{n^2} \right)^2 + \dots \right] - \frac{1}{32} \hat{\gamma}^2 \frac{\lambda m^2}{n} \left[ 1 - \frac{3}{16} \frac{\lambda m^2}{n^2} + \dots \right],
\end{align}
in perfect agreement with the classical expression (\ref{class2}) for the energy if the oscillation number is quantized as $n$. This is expected because we are quantizing in the deformed sphere. It should be noticed that we could have used as the unperturbed Hamiltonian 
the first two terms of (\ref{ham2}) which come together from the Laplacian on the deformed sphere but the result is the same. 

We can now include the effects of the full ten dimensional deformed spacetime on $\theta$ by taking $\sqrt{-g} = \sqrt{G} \sin^3\theta \cos\theta \dots$ in $\Pi_\theta^2$, that is, 
\begin{align}\label{4.21}
	\Pi^2_\theta &= \frac{1}{G \sin^3\theta \cos\theta} \frac{d}{d\theta} \left( G \sin^3\theta \cos\theta \frac{d}{d\theta} \right)  \nonumber \\
	& = \frac{1}{\sin^3\theta \cos\theta} \frac{d}{d\theta} \left( \sin^3\theta \cos\theta \frac{d}{d\theta} \right) - 2 \hat{\gamma}^2 \sin\theta \cos\theta ( 1 - 2 \sin^2\theta ) \frac{d}{d\theta}. 
\end{align}
Taking the first term of (\ref{4.21}) as the unperturbed Hamiltonian in (\ref{HAM}) the normalized eigenfunctions are expressed in terms of Jacobi polynomials as
\begin{align}
\Psi_n(\theta) = 2 \sqrt{n+1} P_n^{(0,1)}( 1 - 2 \cos^2\theta),
\end{align}
with eigenvalue $E^2_{0,n} = 4n(n+2)$. Then $E_{0,n} = 2n$ for large $n$ and we get agreement with the lowest order term of (\ref{class2}) if the oscillation number is quantized as an even integer. Now the wave function is even about $\theta = \pi/2$. 

The first order correction to the energy in the limit of large $n$ is
\begin{align}\label{4.23}
	\delta_1 E^2_n &=  \int_0^{\pi/2} d\theta \, \sin^3\theta \cos\theta \, \Psi^\star_n \left[ 2 \hat{\gamma}^2 \sin\theta \cos\theta ( 1 - 2 \sin^2\theta ) \frac{d}{d\theta} \right. \nonumber \\
	& \left. + \lambda m^2 \sin^2\theta ( 1 - \hat{\gamma}^2 \sin^2\theta \cos^2\theta ) \right] \Psi_n \nonumber \\
	&= \frac{1}{2} \lambda m^2 \left( 1 - \frac{1}{8} \hat{\gamma}^2 \right) - \hat{\gamma}^2. 
\end{align}
This is precisely the result (\ref{correc1}) with an extra term which came from the one derivative contribution in (\ref{4.21}). This contribution is at the same order of $1/n$ as the other terms. In the case of the deformed Minkowski spacetime (\ref{E2}) they were of different orders and the corresponding term was of lower order. 
To second order we find
\begin{align}\label{4.24}
	\delta_2 E^2_n = \frac{1}{128} \frac{\lambda^2 m^4}{n^2} \left( 1 - \frac{1}{8} \hat{\gamma}^2 \right) + \frac{1}{32} \hat{\gamma}^2 \frac{\lambda m^2}{n^2},
\end{align}
and again we find an extra term which was absent in the deformed Minkowski spacetime case. 
Notice that the Hamiltonian mixes terms of order $\hat{\gamma}^2$ and $\lambda m^2$ so that both corrections are present in the energy.
Then the energy of highly excited states is 
\begin{align}
	E = 2n \left[ 1 + \frac{1}{4} \frac{\lambda m^2}{(2n)^2} - \frac{1}{64} \left( \frac{\lambda m^2}{(2n)^2} \right)^2 + \dots \right] - \frac{1}{32} \hat{\gamma}^2 \frac{\lambda m^2}{2n} \left( 1 - \frac{3}{16} \frac{\lambda m^2}{(2n)^2} + \dots \right) - \frac{1}{4} \frac{\hat{\gamma}^2}{n} .
\end{align}
The extra contributions coming from the corrections to the energy in (\ref{4.23}) and (\ref{4.24}) gave rise to the last term $\hat{\gamma}^2/(4n)$. Up to this term we have agreement with (\ref{class2}) if the oscillation number is quantized as $2n$.

Finally we can incorporate some contribution from the $AdS$ sector along the lines of \cite{Minahan:2002rc}. We can consider the term $\Pi^2_\rho$ in (\ref{HAM}) and its contribution to the energy. 
We find that for the unperturbed Hamiltonian there is a shift in the energy $E \rightarrow E + 2(N_\rho+1)$ where $N_\rho$ is the quantum number associated to $\rho$. Since we are considering highly excited oscillation states we can ignore $N_\rho$ assuming that the wave function is concentrated at the origin of the $AdS$ allowing us to take $\rho=0$  so that the energy is just shifted by 2. The same result holds when we include the deformation because the structure of the equations in $AdS$ remains intact.

\section{Conclusions}

We have considered pulsating strings in deformed Minkowski spacetime and $AdS_5 \times S^5_{\hat{\gamma}}$ for small deformation. At the classical level we have found the relation between the oscillation number and the energy. For $AdS_5 \times S^5_{\hat{\gamma}}$ the oscillation number can be expressed in terms of elliptic functions. We have different expansions for low and large energy strings. Since the oscillation number is an adiabatic invariant it can be used to probe the semi-classical regime when $N$ is very large. We then performed the quantization for highly excited string states. In the deformed Minkowski spacetime the radial quantization lead to wave functions which are generalized Laguerre polynomials. The energy can then be computed in perturbation theory and corresponds to the expression of the classical energy if the oscillation number is quantized as an even number. For the $AdS_5 \times S^5_{\hat{\gamma}}$ we have found that the oscillation number can be expressed in terms of elliptic functions which have different forms depending on whether we consider small or large energy. The quantization in the low energy regime, corresponding to short strings, has been performed in \cite{Beccaria:2010zn}. We have quantized the oscillatory motion of the string in the large energy case and found that wave functions are Jacobi polynomials. The energy was computed in perturbation theory and we found one extra term which was not predicted by the classical relation between the energy and the oscillation number. The oscillation number has to be quantized as an even number and the extra term depends exclusively on the deformation and not on the string tension or the winding number. It would be interesting to study the stability of this class of strings along the lines of \cite{Frolov:2005ty}.

It is known that the gauge theory operator dual to pulsating strings in $AdS_5 \times S^5$ is composed of non holomorphic products of the complex scalar fields and that there is a precise match between the energy and the anomalous dimension \cite{Engquist:2003rn,Kruczenski:2004cn,Beisert:2003xu,Minahan:2004ds}. It would be interesting to extend these results to the deformed case using the results of \cite{Gromov:2010dy}. 

As a last remark we note that for a rotating string in $AdS_5 \times S^5_{\hat{\gamma}}$ with angular momentum $J$ and winding number $m$ it is known that its energy can be obtained from the $AdS_5 \times S^5$ energy simply by replacing $|m|$ by $|m + \frac{1}{2} \hat{\gamma} J|$ \cite{Frolov:2005ty}. This is also true for the fluctuations around the classical solution. For pulsating strings it is easily seen that no such property is present.

\acknowledgments

The work of S.G. was supported by Capes and CNPq. The work of V.O.R. is supported by
CNPq grant 304116/2010-6 and FAPESP grant 2008/05343-5.

\end{document}